\documentclass[twocolumn,showpacs,superscriptaddress,pra,floatfix]{revtex4-1}
\usepackage{amsmath}
\usepackage{amssymb}
\usepackage{latexsym}
\usepackage{array}
\usepackage{amsfonts}
\usepackage{mathrsfs}
\usepackage{color}
\usepackage{hyperref}
\usepackage{graphicx}
\usepackage{subfig}

\begin{document}
\title{Multisetting Greenberger-Horne-Zeilinger theorem}


\author{Junghee Ryu}
\email{rjhui@hanyang.ac.kr}
\affiliation{Institute of Theoretical Physics and Astrophysics, University of Gda\'{n}sk, 80-952 Gda\'{n}sk, Poland}

\author{Changhyoup Lee}
\affiliation{Centre for Quantum Technologies, National University of Singapore, 3 Science Drive 2, 117543 Singapore, Singapore}

\author{Zhi Yin}
\affiliation{Institute of Theoretical Physics and Astrophysics, University of Gda\'{n}sk, 80-952 Gda\'{n}sk, Poland}

\author{Ramij Rahaman}
\affiliation{Department of Mathematics, University of Allahabad,
Allahabad-211002, U.P., India}
\affiliation{Institute of Theoretical Physics and Astrophysics, University of Gda\'{n}sk, 80-952 Gda\'{n}sk, Poland}

\author{Dimitris G. Angelakis}
\affiliation{Centre for Quantum Technologies, National University of Singapore, 3 Science Drive 2, 117543 Singapore, Singapore}
\affiliation{School of Electronic and Computer Engineering, Technical University of Crete, Chania, 73100 Crete, Greece}

\author{Jinhyoung Lee}
\email{hyoung@hanyang.ac.kr}
\affiliation{Department of Physics, Hanyang University, Seoul 133-791, Korea}

\author{Marek \.{Z}ukowski}
\email{marek.zukowski@univie.ac.at}
\affiliation{Institute of Theoretical Physics and Astrophysics, University of Gda\'{n}sk, 80-952 Gda\'{n}sk, Poland}

\begin{abstract}
We present a generalized Greenberger-Horne-Zeilinger (GHZ) theorem, which involves more than two local measurement settings for some parties, and cannot be reduced to one with less settings. Our results hold for an odd number of parties. We use a set of observables, which are incompatible but share a common eigenstate, here a GHZ state. Such observables are called concurrent. The idea is illustrated with an example of a three-qutrit system and then generalized to systems of higher dimensions, and more parties. The GHZ paradoxes can lead to, e.g., secret sharing protocols.
\end{abstract}
\maketitle

\newcommand{\bra}[1]{\langle #1\vert}
\newcommand{\ket}[1]{\vert #1\rangle}
\newcommand{\abs}[1]{\vert#1\vert}
\newcommand{\avg}[1]{\langle#1\rangle}
\newcommand{\braket}[2]{\langle{#1}|{#2}\rangle}
\newcommand{\commute}[2]{\left[{#1},{#2}\right]}
\newenvironment{rcases}
  {\left.\begin{aligned}}
  {\end{aligned}\right\rbrace}

Einstein, Podolsky and Rosen (EPR) \cite{EPR35} wanted to show that the mathematical formalism of quantum mechanics, though consistent and giving correct predictions,  is incomplete. This gave birth to the fundamental debate ``\emph{Can quantum mechanical description of physical reality be considered complete}?'' Theories compatible with the EPR's ideas are called ``local-realistic (LR) theories." The basic notion introduced by EPR was the one of (local) elements of reality. These are values of possible measurements of an observable, which can be, in principle, determined without actually performing a measurement. They argued that, if one considers perfect correlations of certain measurements  on pairs of entangled systems (say $A$ and $B$),  such values are an inevitable consequence. Perfect correlations are such that a result on the $A$ side of the experiment, precisely determines the value of another measurement on the $B$ side. If systems are far enough from each other, this means that one can determine the  value at $B$ without any disturbance. Thus, it must be defined even without the act of measurement on side $A$. As elements of reality are missing from the quantum formalism, it is incomplete. 

Bell countered the EPR interpretation of quantum theory \cite{Bell64}. He showed that quantum correlations,  for two spins $1/2$ in a singlet state, cannot be reproduced by LR theories, as they violate an inequality satisfied by all LR predictions. For other early developments, see, e.g., the review of Clauser and Shimony~\cite{Clauser78}. Experiments, up to some loopholes, invalidate local realistic models of quantum states (for a current review see~\cite{Pan12,*Brunner13arxiv}). The research moved to finding new nonclassical phenomena for systems more complicated than two spins $1/2$ (see, e.g.,  \cite{Mermin90c,*Werner01,*Zukowski02,*Collins02,*Laskowski04,*Son06,*James10,*Grandjean12}). As applications of the highly nonclassical correlations were found, this sparked off
 quantum information science. Now we have quantum key distribution, quantum teleportation, generators of truly random numbers, etc.,~\cite{Ekert91,Horodecki96,Cleve97,*Hillery99,*Kempe99,*Scarani01,*Barrett05,*Acin07,Pironio10}.

Greenberger, Horne, and Zeilinger (GHZ)~\cite{GHZ89} presented an  {\it ``all-versus-nothing''} conflict between
local realism, in the form of EPR elements of reality and quantum mechanics,
known now as the GHZ paradox. 
 Mermin~\cite{Mermin90a} gave a very simple GHZ contradiction  for predictions for  a three-qubit GHZ state $\ket{\Psi}=\frac{1}{\sqrt{2}} (\ket{000}-\ket{111})$ shared by
three,  distant from each other, observers. They 
perform randomly chosen local measurements of Pauli observables, 
$\hat{\sigma}_{k}^{x}$ or $\hat{\sigma}_{k}^{y}$, where $k=1,2,3$ numbers 
the observers.  Such measurements, in four cases,  form compatible composite observables
$\hat{v}_1=\hat{\sigma}_{1}^{x} \otimes \hat{\sigma}_{2}^{y} \otimes
\hat{\sigma}_{3}^{y}$, $\hat{v}_2=\hat{\sigma}_{1}^{y} \otimes
\hat{\sigma}_{2}^{x} \otimes \hat{\sigma}_{3}^{y}$,
$\hat{v}_3=\hat{\sigma}_{1}^{y} \otimes \hat{\sigma}_{2}^{y} \otimes
\hat{\sigma}_{3}^{x}$, and $\hat{v}_4=-\hat{\sigma}_{1}^{x} \otimes
\hat{\sigma}_{2}^{x} \otimes \hat{\sigma}_{3}^{x}$, as $\ket{\Psi}$ is a
common eigenstate of these four operators.   We have perfect correlations: $\bra{\Psi} \hat{v}_l
\ket{\Psi}=1$ for all $l$.  With a given perfect correlation, according to EPR, one can define the elements of reality. One can predict with
certainty and without any disturbance  the remote  measurement outcome of
the third local observable, once the other two local results of
measurements are known.   {
The elements of reality} $m_{k}^{x(y)}= \pm 1$ related to $\hat{\sigma}_{k}^{x(y)}$, to reflect the correlations, must satisfy the
following relations: $m_{1}^{x} m_{2}^{y} m_{3}^{y}=1$, $m_{1}^{y}
m_{2}^{x} m_{3}^{y}=1$, $m_{1}^{y} m_{2}^{y} m_{3}^{x}=1$, and $m_{1}^{x}
m_{2}^{x} m_{3}^{x}=-1$. However, all these multiplied  imply 
$1 = -1$.  The elements of
reality are thrown overboard, and so is any attempt to deduce local
realism from quantum perfect correlations.

GHZ-type all-versus-nothing theorems, unlike Bell's, do not use statistical inequalities. The GHZ theorem was generalized to more complex systems, such as multipartite and/or high-dimensional ones ~\cite{Zukowski99,Cerf02a,Lee06,Tang13,Ryu13}, and includes {\em ``all-versus-something"} GHZ-type contradictions~\cite{Kaszlikowski02}. Still, there are many unstudied cases, including some simple ones, e.g., a genuine four-qubit GHZ theorem with two settings per qubit and a genuine three-qutrit GHZ theorem with multisettings. Recently, Tang {\it et al.} investigated a four-qubit GHZ contradiction with many measurement settings~\cite{Tang13a}.  

Here, we  show that one can have  GHZ contradictions for three or more qudit systems, which involve more than two settings for some of the observers, and cannot be reduced to ones with less settings.  To this end, we employ a set of so-called {\em concurrent} composite observables. They do {\em not} commute; however, they still share a {\em common eigenstate}, here a generalized GHZ state~\cite{Lee06,Ryu13}. Our local observables are obtained by using unitary operations involving phase shifters. They can be realized with multiport beam splitters (see Refs.~\cite{ZZH,Zukowski99,Lee06,Lee05}). GHZ-type contradictions find applications in various quantum information tasks, especially in variations of secret sharing protocols~\cite{Zukowski98}, and reduction of communication complexity~\cite{Brukner04}. Thus, an irreducible class of such ``paradoxes" allows a different class of such quantum applications.

We illustrate the concurrent observable idea, and construct a GHZ theorem for three $3d$-dimensional systems involving three settings ($d$ is a positive integer). We extend it to three systems of  dimension $D=Md$, and for $M$  settings. Finally, we generalize our GHZ theorem to an odd number of parties, $N$, and systems of dimension $D=Nd$. If $N$ is prime the paradoxes cannot be reduced to less settings.

With concurrent observables  Lee {\it et al.}~\cite{Lee06} proposed a  GHZ theorem for an $N$-qudit system where $N$ is odd and $D$ is even. Recently, Ryu {\it et al.}~\cite{Ryu13} gave a generalized version of a GHZ theorem for an $N$-qudit system. They also used  concurrent observables. This allowed one to extend the GHZ theorem beyond the results of  Refs.~\cite{Zukowski99,Kaszlikowski02,Cerf02a}. 
Here we follow an extension of the approach introduced in Ref.~\cite{Ryu13}. We  use unitary observables, such as $\hat{M}=\sum_{n=0}^{D-1}\omega^n \left|n\right>_{m}\left<n\right|$, where $\omega=\exp(2 \pi i/D)$. Operator $\hat{M}$  can be paired  with a Hermitian one $\hat{H}$ by setting $\hat{M}=\exp(i \hat{H})$. The complex eigenvalues of $\hat{M}$ can be associated with specific measurement results. Such a unitary representation is very handy~\cite{Cerf02a, Lee06}.

{\em Construction of a GHZ theorem for a three-qudit system.}  Consider the following three-qudit GHZ state:
\begin{equation}
\ket{\psi}=\frac{1}{\sqrt{D}}\sum_{n=0}^{D-1} \ket{n,n,n},
\label{eq:3_ghz}
\end{equation}
where $\{\ket{n}\}_{n=0}^{D-1}$ denotes the basis states for a qudit system. Take a composite operator $\hat{V}\equiv\hat{X} \otimes \hat{X} \otimes \hat{X}$, where $\hat{X}=\sum_{n=0}^{D-1} \ket{n}\bra{n+1\mod D}$. One has $\hat{V}\ket{\psi}=\ket{\psi}$. We  construct concurrent observables by applying unitary operations $\hat{U}= \hat{P}_{1} \otimes \hat{P}_{2}\otimes \hat{P}_{3}$ on $\hat{V}$ with phase shift operators given by $\hat{P}_k=\sum_{n=0}^{D-1} \omega^{f_k(n)} \ket{n}\bra{n}$, where $\omega = \exp(2\pi i/D)$. For each $n$, if ``phases" $f_{k}(n)$ satisfy the condition
\begin{equation}
f_{1} (n) +f_{2} (n)+f_{3} (n) = \gamma n,
\label{eq:invariant_condition}
\end{equation}
for some integer $\gamma$,  the unitarily transformed observables $\hat{V}_{U}=\hat{U}\hat{V}\hat{U}^{\dagger}$ are concurrent and the GHZ state (\ref{eq:3_ghz}) is their common eigenstate with eigenvalue $\omega^{-\gamma}$. For the phases satisfying $f_k (n) = \alpha_k n$, with a rational number $\alpha_k$, each local observable $\hat{X}(\alpha)=\hat{P} \hat{X} \hat{P}^{\dagger}$ is given by
\begin{equation}
  \hat{X}(\alpha)=\omega^{-\alpha} \left( \sum_{n=0}^{D-2} \ket{n} \bra{n+1} + \omega^{ \alpha D} \ket{D-1} \bra{0} \right).
  \label{eq:obs_y}
\end{equation}
Two local observables $\hat{X}(\alpha)$ and $\hat{X}(\beta)$ are inequivalent unless $\alpha - \beta$ is an integer~\cite{Ryu13}.

Consider a three-qudit GHZ state (\ref{eq:3_ghz}) shared by three distant parties, one qudit each. Assume that the first two parties can measure their  qudits using one of three observables $(\hat{X},\hat{Y},\hat{Z})$, whereas the third party chooses between  two,  say $\hat{X}$ and $\hat{Y}$. The outcomes of each measurement are $\omega^{l}$, where $l=0,1,...,D-1$. 

Take  concurrent observables obtainable by operations $\hat{U}$ with phases $f_k (n) \in \{0, n/3, 2n/3 \}$. Set  $\hat{X}=\hat{X}(0)$, $\hat{Y}=\hat{X}(1/3)$, and $\hat{Z}=\hat{X}(2/3)$. The GHZ state (\ref{eq:3_ghz}) is an eigenstate of the following  observables (the last column gives the associated eigenvalues of the observables):
\begin{equation}
\begin{tabular}{l ll l | c}
$\mathcal{O}_1$&=~$\hat{X}$ & $\otimes$ $\hat{X}$ & $\otimes$ $\hat{X}$ ~&   ~$1$, \\
$\mathcal{O}_2$&=~$\hat{Y}$ & $\otimes$ $\hat{Z}$ & $\otimes$ $\hat{X}$ ~&   ~$\omega^{-1}$,  \\
$\mathcal{O}_3$&=~$\hat{Z}$ & $\otimes$ $\hat{Y}$ & $\otimes$ $\hat{X}$ ~&   ~$\omega^{-1}$,  \\
$\mathcal{O}_4$&=~$\hat{X}$ & $\otimes$ $\hat{Z}$ & $\otimes$ $\hat{Y}$ ~&   ~$\omega^{-1}$,  \\
$\mathcal{O}_5$&=~$\hat{Y}$ & $\otimes$ $\hat{Y}$ & $\otimes$ $\hat{Y}$ ~&   ~$\omega^{-1}$,  \\
$\mathcal{O}_6$&=~$\hat{Z}$ & $\otimes$ $\hat{X}$ & $\otimes$ $\hat{Y}$ ~&   ~$\omega^{-1}$.  \\
\end{tabular}
\label{eq:3ghz_qm}
\end{equation}
This holds because the phases $f_k(n)$ of each  observable $\mathcal{O}_i$ satisfy condition~(\ref{eq:invariant_condition}).

Local realistic  outcomes of the measurements, to reproduce correlations~(\ref{eq:3ghz_qm}), must
obey \begin{eqnarray}
\omega^{x_1}\omega^{x_2}\omega^{x_3}&=&1, \nonumber \\
\omega^{y_1}\omega^{z_2}\omega^{x_3}&=&\omega^{-1}, \nonumber \\
\omega^{z_1}\omega^{y_2}\omega^{x_3}&=&\omega^{-1}, \nonumber \\
\omega^{x_1}\omega^{z_2}\omega^{y_3}&=&\omega^{-1}, \nonumber \\
\omega^{y_1}\omega^{y_2}\omega^{y_3}&=&\omega^{-1}, \nonumber \\
\omega^{z_1}\omega^{x_2}\omega^{y_3}&=&\omega^{-1},
\label{eq:3ghz_lr}
\end{eqnarray}
where $\omega^{m_k}$ is the $k$th party's outcome, for $m=x,y,z$. These LR relations give a GHZ-type contradiction with  quantum mechanics:

\begin{enumerate}
\item
Divide the six LR predictions~(\ref{eq:3ghz_lr}) into two subsets of correlations:
\begin{equation}
\{\mathcal{A}_1 ,\mathcal{A}_2 ,\mathcal{A}_3 \}= \{ \omega^{x_1}\omega^{x_2}\omega^{x_3},\omega^{y_1}\omega^{z_2}\omega^{x_3},\omega^{z_1}\omega^{y_2}\omega^{x_3}\} \nonumber
\end{equation}
and
\begin{equation}
\{ \mathcal{B}_1 ,\mathcal{B}_2 ,\mathcal{B}_3 \} = \{ \omega^{x_1}\omega^{z_2}\omega^{y_3},
\omega^{y_1}\omega^{y_2}\omega^{y_3},
\omega^{z_1}\omega^{x_2}\omega^{y_3}\}. \nonumber
\end{equation}

\item
Multiply these LR predictions to get $\prod_{k=1}^{3} \mathcal{A}_k\overline{\mathcal{B}_k}$, where $\overline{x}$ is the complex conjugate of $x$.

\item
This gives $\omega^{3(x_3 - y_3)-1} = 1$, which we call the LR condition. Since $\omega=\exp(2 \pi i /D)$, if $D=3d$, where $d$ is an integer,  there is no integer solution of $
\xi$ for the equation $3\xi-1 \equiv 0 \mod D$. Hence, we have a GHZ-type contradiction.
\end{enumerate}

It is worth noting, that  the third party uses just two settings, while the first two parties use three. Above, the third party chooses between $\hat{X}$ and $\hat{Y}$, but due to the symmetry, one has a similar contradiction for the other two cases of  $\hat{X}$ and $\hat{Z}$, or $\hat{Y}$ and $\hat{Z}$. In general, the LR condition leads to $\omega^{3\xi-\eta_3}=1$, where $\xi$ gives the difference between the LR values of two observables selected by the third party, and $\eta_3=3\abs{\alpha_3 - \alpha'_{3}}$. Here, $\alpha_3$ and $\alpha'_3$ are rational numbers associated with the phases defining the two observables of the third party, respectively. As $\alpha_3, \alpha'_3\in \{0, 1/3, 2/3 \}$, $\eta_3$ is always a positive integer, smaller than the number of local observables $M$, here $M=3$. Since $\omega=\exp(2 \pi i /D)$ and $D=3d$, the LR condition is equivalent to the existence of an integer solution for $\xi$ satisfying   $3\xi-\eta_3 \equiv 0 \mod D$. This is impossible.  Hence, we get a GHZ-type contradiction.


One cannot arrive at a GHZ contradiction by using a {\em subset} of the  correlations which involves just two settings, for the first party or the second party. The correlation conditions  are equivalent to the following six equations (in a modulo 3 algebra) involving eight variables:
\begin{eqnarray}\label{ALGEBRA}
(x_1)& ~ + ~ x_2 ~+& ~ (x_3)=a,  \nonumber \\
 y_1& ~ + ~ z_2 ~+& ~ x_3 =2,\hspace{1cm}(*)\nonumber \\
 z_1& ~ + ~ (y_2) +& ~ (x_3) =2,\nonumber \\
 (x_1)& ~ + ~ (z_2) +& ~ y_3 =2,\nonumber \\
 y_1& ~ + ~ y_2 ~+& ~ y_3 =2, \hspace{1cm}(*)\nonumber \\
 z_1& ~ + ~ x_2 ~+& ~ y_3 =2,  \\  \nonumber
\end{eqnarray}
where in the case of our GHZ contradiction $a=0$. The brackets denote variables which will change their values during our proof, below. Only if  we put $a=2$, are the conditions consistent; that is, they can be satisfied with variables of integer values (just put $x_3=y_3=2$, and for all other variables put $0$). However, for our case of $a=0$,  this is impossible, as shown above by our three-step method. Now, if we remove one of the settings, e.g., the $\hat{Y}$ of the first  party, this removes the second and fifth equation from the set, marked by $(*)$.  As one has a proper integer solution for the full set of equations with $a=2$, this solution also satisfies the four-equation subset, with $a=2$ in the first equation. Nevertheless, there  is also a set of integer values for all variables which fit the {\em four} equations in the case of $a=0$. One can do the following: take the values of the variables  for the case $a=2$; next,  lower the values of $x_1$ and $x_3$ by one, and compensate this  in the third and fourth equations by increasing by one $y_2$ and $z_2$ (these two variables  appear only in the third and fourth equations, as marked by the parentheses).  Such operations change the value of the left-hand side (LHS) of the first equation only, and thus the right-hand side (RHS) must now be $a=0$. All new values of the variables are still integers; the four equations are satisfied by them, for $a=0$. Thus, there is no possibility of finding a GHZ contradiction using only these {\em four} equations, as a set of integer valued variables satisfies them. {\em Ipso facto}, our  GHZ argument necessarily involves all six correlations. Obviously,  the above argument still holds when one tries to remove any other setting of the first or second party.

This does not preclude the possibility that for our system and the state, for {\em different} sets of settings, one may get a GHZ contradiction involving fewer settings. But this is not our aim. Irreducibly multisetting paradoxes lead to quantum solutions of secret sharing and reduction of communication complexity problems, which cannot be related to two settings-per-party GHZ paradoxes. 

The question of a minimal set of measurements for the given situation is an interesting open problem. Recently, in Ref.~\cite{Lawrence13a} Lawrence addressed the GHZ contradiction for $N$ parties and $D$ dimensional systems by using fewer measurement settings, but the set of composite observables he used is different from ours. 



{\em Arbitrary  number of settings, $M>3$.} To this end, consider the phases $f_k (n)=\alpha_k n \in \{0, n/M, \dots, (M-1)n/M \}$. Following a similar procedure as above one can construct $2M$ concurrent composite observables, whose common eigenstate is ~(\ref{eq:3_ghz}). The first two parties measure $M$ different observables, while the third party can measure arbitrary two observables from the set of $M$ [see Eq.~(\ref{eq:3ghz_qm})]. Following the method,  divide these $2M$ concurrent observables into two sets $\mathcal{A}$ and $\mathcal{B}$, 
and produce  $\prod_{k=1}^{M} \mathcal{A}_k\overline{\mathcal{B}_k}$. 
The LR condition is in the form of $\omega^{M\xi - \eta_M}=1$. Here $\eta_M=M\abs{\alpha_3 - \alpha'_3}$ with $\alpha_3, \alpha'_3 \in \{0, 1/M, \dots, (M-1)/M \}$. If $D=Md$, there is no integer solution for  $\xi$ satisfying  $M\xi-\eta_M \equiv 0 \mod D$, because $\eta_M$ is a nonzero integer,  and  $\eta_M<M$. We have a $M$-setting tripartite $(Md)$-dimensional GHZ contradiction.

In the case of a {\em prime} $M>1$,
 to show the irreducibility of the number of settings required for such GHZ contradiction, one can use a  generalization of the  argumentation using Eqs.~(\ref{ALGEBRA}). One can build  a set of equations of a similar kind for the perfect correlations.  Once again, one is able to show that a {\em subset} of the equations involving less settings cannot lead a GHZ contradiction. Whenever we remove one of the settings for the first or second party, in the remaining subset of equations there are always ``lone" variables,  which  appear in just one equation, just like $y_2$ and $z_2$ in ({\ref{ALGEBRA}}). This is so,  because  the method to construct our multisetting GHZ contradictions uses cyclic permutations to get the sequences of composite observables. Again one can find such value of the RHS of the first equation such that variables of  integer values satisfy the full set. By proper changes of the values of the variables in the first equation, and compensating changes of the ``lone" variables, one can always produce an integer solution  for the subset of equations, with RHSs as in the case of the contradiction. For multipartite cases, this argument still works for a prime number of settings, as we also will use the cyclic permutations to construct the multipartite GHZ contradictions (we will explain this below). 

If $M$ is a nonprime number, one can find a GHZ contradiction with a {\em subset} of  the perfect correlations. For example, take a four-setting GHZ contradiction using  local observables defined by  phases  $0, 1/4, 2/4, 3/4$. In a subset of perfect correlations, in which there  are only  observables with the ``phases" $0$ and $2/4$, there are no ``lone'' variables. This subset of correlations  can be found in Ref.~\cite{Lee06} to give a GHZ contradiction. Generally, the number of measurement settings required for a contradiction can be reduced to one of prime divisors of $M$.


{\em More complex systems.}
One can generalize the method to the  $N$-partite $D$-dimensional case with  $N$ measurement settings for each party, provided $N$ is an  odd integer and $D=Nd$. This requires $2N$ concurrent composite observables. The first $N-1$ parties choose between $N$ different local settings and the last party between two, out of the set of  $N$. Consider $N$ qudits prepared in a  GHZ state
\begin{equation}
\ket{\psi}=\frac{1}{\sqrt{D}} \sum_{n=0}^{D-1} \bigotimes_{k=1}^{N} \ket{n}_k.
\label{eq:ND_ghz_state}
\end{equation}
Each party chooses a local observable from $\{\hat{X}(\frac{l}{N})\}_{l=0}^{N-1}$ given in Eq.~(\ref{eq:obs_y}) with the phases $f (n)=\frac{l}{N} n$. For simplicity, assume that the $N$th party selects between two measurements $\hat{X}(0)$ and $\hat{X}(1/N)$ (later we discuss the other cases). Then, the corresponding $2N$ concurrent composite observables are given by
\begin{eqnarray}\label{eq:Nghz_qmA}
\hat{\mathcal{A}}_r=&\hat{X}&(\frac{r\ominus 1}{N})\hat{X}(\frac{r\ominus 2}{N})\cdots \nonumber\\
&\otimes& \hat{X}(\frac{r\ominus (N\ominus 2)}{N})\hat{X}(\frac{t_r}{N})\hat{X}(0)
\end{eqnarray}
and
\begin{eqnarray}\label{eq:Nghz_qmB}
\hat{\mathcal{B}}_r=&\hat{X}&(\frac{r\ominus 1}{N})\hat{X}(\frac{r\ominus 2}{N})\cdots \nonumber\\  
 &\otimes& \hat{X}(\frac{r\ominus (N\ominus 2)}{N})\hat{X}(\frac{t_r\ominus 1}{N})\hat{X}(\frac{1}{N})
\end{eqnarray}
for $r=1,2,\dots,N$ with $t_r\equiv 2r \oplus 1$. Tensor product symbols are omitted, except for the line break. Here, ``$\oplus$" (``$\ominus$") denotes the addition (subtraction) under mod $N$, and $t_r \not\equiv t_j$ for $r\not\equiv j$ mod $N$.

All  composite observables $\{\hat{\mathcal{A}}_r\}_{r=1}^N$ and $\{
\hat{\mathcal{B}}_r\}_{r=1}^N$ satisfy their corresponding invariance condition~(\ref{eq:invariant_condition}). Therefore, the  state~(\ref{eq:ND_ghz_state}) is their common eigenstate:
\begin{equation}
\hat{\mathcal{A}}_r\ket{\Psi}=\omega^{-\gamma_r}\ket{\Psi} \mbox{~~ and ~~ }\hat{\mathcal{B}}_r\ket{\Psi}=\omega^{-\gamma'_r}\ket{\Psi},
\end{equation}
where $\gamma_r=\frac{1}{N}\left(\sum_{k=1}^{N-2} r\ominus k+t_r\right)$ and $\gamma'_r=\frac{1}{N}[\sum_{k=1}^{N-2} r\ominus k+(t_r\ominus 1)+1]$. Note that $\gamma_r$ and $\gamma'_r$ are, in general, nonzero integers less than $N$.

Following the method, with the products $\prod_{r=1}^{N} \mathcal{A}_r\overline{\mathcal{B}_r}$ we obtain the LR condition $\omega^{N(x_{N}^{0} - x_{N}^{1/N})-1}=1$. Here, $\omega^{x_{N}^{0}}$ $(\omega^{x_{N}^{1/N}})$ denotes the LR value of the $N$th party's measurement $\hat{X}(0)$ $[\hat{X}(\frac{1}{N})]$.
However,  $\omega=\exp(2\pi i /D)$ and $D=Nd$ (for an integer $d$). 
Thus, the LR condition cannot hold: if $N\xi-1 \equiv 0 \mod D$ holds, then  $\xi$ cannot be an integer. 
If the $N$th party chooses two other measurements, say  $\hat{X}(\alpha)$ and $\hat{X}(\alpha')$, then the LR condition leads to $\omega^{N(x_{N}^{\alpha} - x_{N}^{\alpha'})-\eta_N}=1$, where $\eta_N = N \abs{\alpha_{N}-\alpha'_{N}}$ with $\alpha_{N},\alpha'_{N} \in \{0, 1/N, 2/N, \dots, (N-1)/N \}$. Again, $\eta_N$ is a positive integer smaller than $N$. Like earlier, in this case also the value of $x_{N}^{\alpha} - x_{N}^{\alpha'}$ cannot be an integer, if equation  $N(x_{N}^{\alpha} - x_{N}^{\alpha'})-\eta_N\equiv 0 \mod D$ is to hold. Thus, we have  a  GHZ contradiction.

The final question  is whether the above GHZ contradictions cannot be reduced to ones involving lower dimensions, or less particles. 
In 2002, Cerf {\it et al.} \cite{Cerf02a} introduced the genuineness criterion for  a generalized GHZ theorem. A  GHZ argument is called genuine, if one cannot obtain another GHZ-type contradiction from this argument by reducing the number of parties or the dimension of any subsystem. 
Our $N$-partite GHZ arguments are genuine. They are constructed using a set of composite observables following cyclic permutations [see the $2N$ concurrent observables given in Eqs.~(\ref{eq:Nghz_qmA}) and (\ref{eq:Nghz_qmB})]. If we eliminate one of the parties, we are unable to show a GHZ contradiction with the remaining observables. The $N$-partite GHZ state is no longer their common eigenstate.

The genuine $D$ dimensionality of our argument is reflected by the fact that the operators which we use  are undecomposable into a direct sum of  subdimensional ones ~\cite{Lee06,Ryu13}. Assume the contrary: for one of the parties one can simultaneously block diagonalize all   $\hat{X}(\alpha)$ operators which are involved in the argument. Thus, for each $\hat{X}(\alpha)$  we have at least one splitting into a direct sum $\hat{X}(\alpha)=\hat{X}(\alpha)_{D-K}\oplus\hat{X}(\alpha)_K$, where $D-K$ and $K$ are the dimensions of the subspaces, which via the direct sum reproduce the original full space. Of course, to reduce the dimension of our argument one has to have the same type of direct sum splitting of all observables [for each $\alpha$ the operator $\hat{X}(\alpha)_{D-K}$ acts on the same $D-K$ dimensions, etc.]. In such circumstances for any two noncommuting operators $\hat{X}(\alpha)$  and $\hat{X}(\alpha')$, one can find pairs of eigenvectors---$|e\rangle_{\alpha}$ for the first one and $|e'\rangle_{\alpha'}$ for the second one---such that  $_{\alpha'}\langle e'|e\rangle_{\alpha}=0$. Simply $\ket{e}_{\alpha}$, e.g., may be in  the  $(D-K)$-dimensional subspace, while $\ket{e'}_{\alpha'}$ may be in the $K$-dimensional one. But this is not so for the operators involved in our argument.  The eigenvectors of the local observable $\hat{X}(\alpha) $ are  $\ket{n}_{\alpha } = \frac{1}{\sqrt{D}} \sum_{m=0}^{D-1} \omega^{(n+\alpha) m} \ket{m}$.  This implies that for every $n$ and $m$, one has $| _{\alpha}\langle n \ket{m}_{\alpha'} |^2= \frac{\sin^2 (\pi \xi)}{D^2 \sin^2 \left[(\pi/D) \xi \right]} > 0$,
where $\xi=m-n+\alpha'-\alpha$. Thus the simultaneous block diagonalization is impossible. With the used observables the argument is irreducibly $D$ dimensional.


In summary, we have discussed the problem of generalization of a GHZ theorem, and present here the results for an odd number of parties, $N$.  To construct our theorem we adopt the concurrent observable approach \cite{Lee06,Ryu13}. The invariance condition~(\ref{eq:invariant_condition}) guarantees a common eigenstate for a set of observables, even if they are incompatible. For a prime $N$, we show an irreducible all-versus-nothing GHZ contradiction 
for  an $N$-partite $D=(Nd)$-dimensional system involving $N$ local measurement settings. The preliminary version of this work can be found in~\cite{Ryu13arxiv}.

{\em Acknowledgements}. We thank M. Wie\'{s}niak, T. Paterek, and J. Jae for comments. This work is supported by the Foundation for Polish Science TEAM project cofinanced by the EU European Regional Development Fund, and by a NCBiR-CHIST-ERA Project QUASAR. R.R acknowledges support by an ERC grant QOLAPS (Grant No. 291348). C.L. and D.A. are supported by the National Research Foundation and Ministry of Education (partly through the Tier 3 Grant ``Random numbers from quantum processes"), Singapore. J.L. is supported by the National Research Foundation of Korea (NRF) grant funded by the Korea government (MEST) (Grants No. 2010-0015059 and No. 2010-0018295).


\begin{thebibliography}{99}

\bibitem{EPR35} A. Einstein, B. Podolsky, and N. Rosen, Phys. Rev. {\bf 47}, 777 (1935).
\bibitem{Bell64} J. S. Bell, Physics {\bf 1}, 195 (1964).
\bibitem{Clauser78} J. Clauser and A. Shimony, Rep. Prog. Phys. {\bf 41}, 1881 (1978).
\bibitem{Pan12} J.-W. Pan, Z.-B. Chen, C.-Y. Lu, H. Weinfurter, A. Zeilinger, and M. \.{Z}ukowski, Rev. Mod. Phys. {\bf 84}, 777 (2012).
\bibitem{Brunner13arxiv} N. Brunner, D. Cavalcanti, S. Pironio, V. Scarani, and S. Wehner, arxiv:1303.2849v2
\bibitem{Mermin90c} N. D. Mermin Phys. Rev. Lett. {\bf 65}, 1838 (1990).
\bibitem{Werner01} R. F. Werner and M. M. Wolf, Phys. Rev. A {\bf 64}, 032112 (2001).
\bibitem{Zukowski02} M. \.{Z}ukowski and \v{C}. Brukner, Phys. Rev. Lett. {\bf 88}, 210401 (2002).
\bibitem{Collins02} D. Collins, N. Gisin, N. Linden, S. Massar, and S. Popescu, Phys. Rev. Lett. {\bf 88}, 040404 (2002).
\bibitem{Laskowski04} W. Laskowski, T. Paterek, M. \.{Z}ukowski, and \v{C}. Brukner, Phys. Rev. Lett. {\bf 93}, 200401 (2004).
\bibitem{Son06} W. Son, J. Lee, and M. S. Kim, Phys. Rev. Lett. {\bf 96}, 060406 (2006).
\bibitem{James10} J. Lim, J. Ryu, S. Yoo, C. Lee, J. Bang, and J. Lee, New J. Phys. {\bf 12}, 103012 (2010).
\bibitem{Grandjean12} B. Grandjean, Y.-C. Liang, J.-D. Bancal, N. Brunner, and N. Gisin, Phys. Rev. A {\bf 85}, 052113 (2012).
\bibitem{Ekert91} A. K. Ekert, Phys. Rev. Lett. {\bf 67}, 661 (1991).
\bibitem{Horodecki96} R. Horodecki, M. Horodecki, and P. Horodecki, Phys. Lett. A {\bf 222}, 21 (1996).
\bibitem{Cleve97} R. Cleve and H. Buhrman, Phys. Rev. A {\bf 56}, 1201 (1997).
\bibitem{Hillery99} M. Hillery, V. Bu\v{z}ek, and A. Berthiaume, Phys. Rev. A {\bf 59}, 1829 (1999)
\bibitem{Kempe99} J. Kempe, Phys. Rev. A {\bf 60}, 910 (1999)
\bibitem{Scarani01} V.  Scarani and N. Gisin, Phys. Rev. Lett. {\bf 87}, 117901 (2001).
\bibitem{Barrett05} J. Barrett, L. Hardy, and A. Kent, Phys. Rev. Lett. {\bf 95}, 010503 (2005).
\bibitem{Acin07} A. Ac\'{i}n, N. Brunner, N. Gisin, S. Massar, S. Pironio, and V. Scarani, Phys. Rev. Lett. {\bf 98}, 230501 (2007).
\bibitem{Pironio10} S. Pironio, A. Ac\'{i}n, S. Massar, A. Boyer de la Giroday, D. N. Matsukevich, P. Maunz, S. Olmschenk, D. Hayes, L. Luo, T. A. Manning, and C. Monroe, Nature {\bf 464}, 1021 (2010).
\bibitem{GHZ89} D. M. Greenberger, M. A. Horne, and A. Zeilinger, in {\em Bell's Theorem, Quantum Theory, and Conceptions of the Universe}, edited by M. Kafatos (Kluwer, Dordrecht, 1989).
\bibitem{Mermin90a} N. D. Mermin, Am. J. Phys. {\bf 58}, 731 (1990).

\bibitem{Zukowski99} M. \.{Z}ukowski and D. Kaszlikowski, Phys. Rev. A {\bf 59}, 3200 (1999).
\bibitem{Cerf02a} N. J. Cerf, S. Massar, and S. Pironio, Phys. Rev. Lett. {\bf 89}, 080402 (2002).
\bibitem{Lee06} J. Lee, S.-W. Lee, and M. S. Kim, Phys. Rev. A {\bf 73}, 032316 (2006).
\bibitem{Tang13} W. Tang, S. Yu, C. H. Oh, Phys. Rev. Lett. {\bf 110}, 100403 (2013).
\bibitem{Ryu13} J. Ryu, C. Lee, M. \.{Z}ukowski, and J. Lee, Phys. Rev. A {\bf 88}, 042101 (2013)
\bibitem{Kaszlikowski02} D. Kaszlikowski and M. \.{Z}ukowski, Phys. Rev. A {\bf 66}, 042107 (2002).
\bibitem{Tang13a} W. Tang, S. Yu, and C. H. Oh, arXiv:1303.6740.
\bibitem{ZZH} M \.Zukowski, A. Zeilinger, M. A. Horne, Phys. Rev. A {\bf 55},  2564 (1997)
\bibitem{Lee05} J. Lee and S.-W. Lee, J. Korean Phys. Soc. {\bf 46}, 181 (2005).
\bibitem{Zukowski98} M. \.{Z}ukowski, A. Zeilinger, M. A. Horne, and H. Weinfurter, Acta Phys. Pol. A {\bf 93}, 187 (1998).
\bibitem{Brukner04} \v{C}. Brukner, M. \.{Z}ukowski, J.-W. Pan, and A. Zeilinger, Phys. Rev. Lett. {\bf 92}, 127901 (2004).
\bibitem{Lawrence13a} J. Lawrence, Phys. Rev. A {\bf 89}, 012105 (2014).
\bibitem{Ryu13arxiv} J. Ryu, C. Lee, M. \.{Z}ukowski, and J. Lee, arXiv:1303.7222v1.

\end{thebibliography}

\end{document}